\documentclass[referee]{raa}            % referee version: for submission

\usepackage{graphicx,times}             %for PS/EPS graphics inclusion, new
\usepackage{multirow}
\usepackage{color}
\begin{document}
    \title{Experiment of Diffuse Reflection Laser Ranging to Space Debris and Data Analysis
%\footnotetext{$*$ Supported by the National Natural Science Foundation of China.}
}
%   \subtitle{I. Place Your Subtitle Here}

   \volnopage{Vol.0 (200x) No.0, 000--000}      %%preserved for Editor. DOn't remove!
   \setcounter{page}{1}          %%starting page, preserved for Editor. DOn't remove!
   \author{Hao Sun
      \inst{1,2}
   \and HaiFeng Zhang
      \inst{1}
   \and ZhongPing Zhang
      \inst{1}
   \and Bin Wu
      \inst{1}
   }

   \institute{Shanghai Astronomical Observatory, Chinese Academy of Sciences,
              Shanghai 200030, China; {\it aiying@bao.ac.cn}\\
        \and
             Graduate University of Chinese Academy of Sciences, Beijing 100049, China\\
   }

   \date{Received~~2009 month day; accepted~~2009~~month day}

\abstract{ Space debris has been posing a serious threat to human space activities and is needed to be measured and cataloged. As a new technology of space target surveillance, the measurement accuracy of DRLR (Diffuse Reflection Laser Ranging) is much higher than that of microwave radar and electro-optical measurement. Based on laser ranging data of space debris from DRLR system collected at SHAO (Shanghai Astronomical Observatory) in March-April 2013, the characteristics and precision of the laser ranging data are analyzed and its applications in OD (Orbit Determination) of space debris are discussed in this paper, which is implemented for the first time in China. The experiment indicates that the precision of laser ranging data can reach 39cm-228cm. When the data is sufficient enough (4 arcs of 3 days), the orbit accuracy of space debris can be up to 50m.
}

   \authorrunning{Hao Sun, HaiFeng Zhang, ZhongPing Zhang,Bin Wu }            %author_head in even pages
   \titlerunning{Experiment of Diffuse Reflection Laser Ranging to Space Debris and Data Analysis } % title_head in odd pages
   \maketitle
%________________________________________________ sections below
%
\section{Introduction}           %% first-level sections will be auto-capitalized
\label{sect:intro}

Space debris known as space waste, is a man-made material trapped around the Earth and comes from rocket bodies, upper stages engines, unused satellite, discards of space missions, and collisions between space objects etc(Liu 2004). In recent years, human spaceflight and space launch activities have become more and more frequent and spatial objects in orbit are increased exponentially. The latest news on the Space-Track website shows that there are about 16,000 pieces of debris which is larger than 10 cm. Space debris is threatening to the safety of the spacecrafts in service, and potential collision risk is a serious constraint to the future spacecraft launch, spacecraft-testing and other space activities. With the development of spaceflight technology, there will be more and more space activities in the future. To ensure the safety of the future space activities, sustainable development and utilization of space resources, the new technologies for tracking and cataloging space debris should be developed.

Currently, the routine measurement systems to space debris and cataloging management are mainly the electronic fence, optical or radar equipment. Limited by the precision of the current measurement equipment from several meters to tens of meters, which cannot meet the requirement of precise measurement to some space objects, the new observation technology is needed. As one of the precise observation technology for the space geodesy, single measurement precision of SLR (Satellite Laser Ranging)  to the cooperative targets with corner reflectors could reach sub centimeter level, and will be up to millimeter level (Ye 2000). But for the measurement of non-cooperative targets without corner reflectors, such as space debris, most satellites and unused spacecrafts, the DRLR (Diffuse Reflection Laser Ranging) technology should be used. The method of diffuse reflection laser ranging is similar to the conventional laser ranging, the distance between a ground-based laser station and space targets is obtained by measuring the round-trip propagation time of the laser signal. The main difference is that the way of laser pulse reflection and the cooperative targets make the incoming laser pulse back to the ground station through the corner reflectors, while through the way of diffuse reflection from the surface for uncooperative targets. The laser energy of diffuse reflection is far less than that from cooperative target. But comparing with the conventional laser observation, diffuse reflection laser ranging technology has a wider range of applications, and its measurement precision can reach to 50 - 250cm (e.g.,Ben 2002;Kirchner 2012;Li 2011;Zhang 2012), which is about 1-2 orders of magnitude better than that of electronic fence, optical or radar system. DRLR could provide high-precision observation and it is beneficial to monitor the space debris.

Based on the 60cm aperture laser ranging system and DRLR technology, some experiments about DRLR for tracking the space debris are carried out using 200Hz, 50W power lasers system in March -April 2013 at SHAO. Lots of valuable laser ranging data are obtained. The laser ranging data of the single station from some space debris is processed to realize the orbit determination for space debris and evaluate the accuracy of the DRLR data. Meanwhile, the application of the DRLR data in orbit determination for space debris is discussed in this paper.

Because of the ubiquity of space object collision risk, accurate surveillance of space target such as satellite and debris becomes very important. Institutions have paid great attention to laser ranging technology tracking and monitoring the space debris and non-cooperative target. Telescope with the aperture of 3.5m of the United States Air Force in New Mexico, Starfire, has been collecting the corresponding data. In 1994, Fugate reported that they could measure non-cooperative targets within 1000 km with defuse reflection laser ranging system on the ninth Canberra International Conference on laser ranging. In recent years, a new development trend was noticed that the diffuse reflection laser ranging technology is used to monitor the uncooperative targets. The Australian EOS company started the laser measurement to space debris at the Stromlo station of Australia Canberra from 2000. In 2002 October, Ben Greene, a member of this company published a report entitled "Laser Tracking of Space Debris" at the thirteenth session of the international laser ranging meeting for the first time in Washington. He introduced the progress and measuring results of defuse reflection laser ranging to uncooperative targets. They can detect a size of 15cm space debris within 1250 km with a 76cm aperture telescope and high power laser system at Stromlo station(Ben 2002). In 2004, the aperture of space debris ranging telescope has been up to 1.8m after reconstruction. In addition, the Austria Graz laser ranging station has successfully tracked 85 arcs of 43 space debris targets on December 11, 2011 and May 10, 2012 (Kirchner 2012). The measured distances are from 600 km to 2500 km. The radar cross sections (RCS) is from 0.3 m$^2$ to 15 m$^2$. The average precision (RMS) is about 0.7 m.

SHAO started to develop DRLR technology in 2006. Cooperating with the Eleventh Institute of Chinese Electronic Technology Group Company, the first set of DRLR system with the laser power of 40W@20Hz, 532nm wavelength have been established at SHAO. Since then, it began to track and measure non-cooperative targets. The 3 arcs laser ranging data from 2 rocket debris were obtained with a precision (RMS) of 68$~$83cm(SHAO 2009) in July 2008. This opened the field of development of DRLR technology in China. After upgrading of the measuring system, the 43 arcs of 18 space target (rocket debris and defunct satellite) were successfully obtained in March and April 2013. Yunnan Astronomical Observatory of CAS also developed the technology of DRLR based on the SLR system with the aperture of 1.2 m telescope and the laser ranging data were successfully obtained from the dozens arcs of space debris in 2010-2011(Li 2011).

\section{DRLR system and laser ranging data statistics of space debris}
\label{sect:DRLR}
During past several years SHAO has built space debris laser ranging system with high pulse energy, low repetition rate, lamp pumped solid-state laser. And for the first time in China, our group successfully applies the 50W power diode pumped solid-state laser system with 200Hz and low dark noise APD detector to observe space debris and many passes of laser data are obtained.

For laser ranging to space debris, the high power laser system with good beam quality, stability and pointing accuracy is very important, especially for observing space objects with long distance and small size. The lamp pumped laser system with low repetition rate was used during 2008-2012 at SHAO. For the high power laser system, the best way to meet the power and not to damage the module of laser system is to increase the working frequency and to decrease the pulse energy. In order to test the measuring ability of high power laser at the high repetition rate mode, a set of the semiconductor pumped laser system with the power of 50W and 200Hz working frequency was installed at SHAO at the beginning of 2013.

The high power laser system helps to solve number of laser returns from space debris. Next the keys is to reduce the level of noise detection to make large scale of range gate adjustment and obtain laser returns with high S/N for farther and smaller space debris. The breadboard APD detector is developed by our group based on Compass LTT detector and cooperation with domestic university. Figure 3 shows the APD detector and its main performances. By using the detector, the dark noise can be decreased. But the noise from sky and targets also make some influences on the laser detection. For that, the high efficiency narrow bandwidth spectrum filter is adopted to reduce the level of background noise. The main characteristics of the filter are following:1)Center wavelength:532nm;2)Bandwidth:1nm;3)Efficiency:$>$90\%.

The event timer Model A033-ET for time interval measurement with 10 psec timing precision made by the Riga University, Latvia is used. The control system of laser measurement is the same as the routine kHz SLR system. The tracking error of 60cm telescope is less than 1 arc second to meet the requirement of tracing to space debris targets.

In March and April 2013, DRLR data of 18 targets which have more than 2 arcs is obtained. Those targets are mainly rocket bodies, unused satellites with orbit altitude (perigee altitude) from 400 to 900km. The minimum target size (RCS) is about 1m$^2$. The maximum is about 12 m$^2$. Table 1 shows the detail information of measured targets. The targets which have more than 3 arcs are marked by bold and inclined on the table.
\begin{table}
\begin{center}
\begin{minipage}[]{100mm}
\caption[]{The detail information of the measured targets by DRLR}\label{Tab:publ-works}
\end{minipage}
\setlength{\tabcolsep}{1pt}
\small
%%Please Capitalize the First Letter of Each Notional Word in table's caption

 \begin{tabular}{ccccccc}
  \hline\noalign{\smallskip}
NORADID & Perigee($km$)& Eccentricity & inclination($deg$) & InterId & RCS($m^2$) & Name\\
  \hline\noalign{\smallskip}
  \itshape\bfseries 20453 & 428.41 & 0.03465 & 35.628 & 1990008B & 9.79 & DELTA 2 R/B(1)\\ % new variable
  \itshape\bfseries 28222 & 514.65 & 0.00691 & 97.367 & 2004012C & 12.25 & CZ-2C R/B \\
  23343 & 630.79 & 0.00133 & 98.203 & 1994074B & 8.94	 & SL-16 R/B \\
  20323 & 688.23 & 0.00611 & 97.070 & 1989089B & 8.90	 & DELTA 1 R/B \\
  \itshape\bfseries 11574 & 739.92 & 0.00228 & 74.072 & 1979089B & 5.43	 & SL-8 R/B \\
  24968 & 765.70 & 0.00125 & 86.394 & 1997056D & 3.95	 & IRIDIUM 37 \\
  23705 & 827.30 & 0.00181 & 71.025 & 1995058B & 9.96	 & SL-16 R/B \\
  16182 & 835.66 & 0.00013 & 71.006 & 1985097B & 12.21 & SL-16 R/B \\
  12138 & 395.94 & 0.07381 & 82.960 & 1981003A & 2.98	 & COSMOS 1238 \\
  20362 & 437.42 & 0.01284 & 35.627 & 1989097B & 9.58	 & DELTA 2 R/B(1) \\
  \itshape\bfseries 12465 & 507.94 & 0.00463 & 81.225 & 1981046B & 5.67	 & SL-3 R/B \\
  16496 & 595.89 & 0.00294 & 82.526 & 1986006B & 4.50	 & SL-14 R/B \\
  14820 & 596.89 & 0.00301 & 82.542 & 1984027B & 3.86	 & SL-14 R/B \\
  19275 & 614.55 & 0.00089 & 82.514 & 1988056B & 4.83	 & SL-14 R/B \\
  18765 & 748.98 & 0.00064 & 66.583 & 1985042H & 1.13	 & SL-12 DEB \\
  \itshape\bfseries 17590 & 829.31 & 0.00077 & 71.005 & 1987027B & 9.64	 & SL-16 R/B \\
  24298 & 834.92 & 0.00194 & 70.868 & 1996051B & 8.79	 & SL-16 R/B \\
  \itshape\bfseries 20788 & 868.20 & 0.00196 & 99.013 & 1990081A & 3.05	 & FENGYUN 1B \\
  \noalign{\smallskip}\hline
\end{tabular}
\end{center}
\end{table}

Table 1 indicates that there are 6 targets having more than 3 arcs. These measurements are implemented on between March 2 and March 7, between April 12 and April 16, between April 26 and April 29 in 2013, respectively. Table 2 shows the detail information of laser ranging data from the 6 targets. The longest time of measured arc is about 4 minutes. The mean of interval sampling data (except 28222) is about 100 ms. In other words, about 10 laser ranging data could be collected in per second. The minimum measured distance is 571km and the maximum is 1455 km. The information of the targets and arcs is shown in Table 2.
\begin{table}
\begin{center}
\begin{minipage}[]{100mm}
\caption[]{The detail information of laser ranging data of DRLR}\label{Tab:publ-works}
\end{minipage}
\setlength{\tabcolsep}{1pt}
\small
%%Please Capitalize the First Letter of Each Notional Word in table's caption
 \begin{tabular}{ccccccc}
  \hline\noalign{\smallskip}
NORADID & Start Time($UTC$)& Len. of Arc($s$) & Len.of sampling($s$) & Echoes & Mean Sampling($per sec$) & Range($km$)\\
  \hline\noalign{\smallskip}
  \multirow{4}{*}{20453} & 03-04 11:49:08 & 77 & 72 & 669 & 9 & \multirow{4}{*}{777-1179}\\% new variable
   & 03-05 10:38:06 & 209 & 197 & 1629 & 8 & \\
   & 03-05 12:21:57 & 101 & 99  & 1443 & 15 \\
   & 03-06 11:12:27 & 131 & 105 & 557  & 5 \\
  \hline
  \multirow{3}{*}{28222} & 03-02 21:06:44 & 98 & 89 & 6010 & 68 & \multirow{3}{*}{571-1287}\\% new variable
   & 03-03 21:06:39 & 30  & 30  & 893  & 30 \\
   & 03-04 21:07:55 & 104 & 98  & 3583 & 37 \\
  \hline
  \multirow{3}{*}{11574} & 03-02 21:22:30 & 96 & 81 & 312 & 4 & \multirow{3}{*}{855-1317}\\% new variable
  & 03-03 20:40:30 & 221 & 198 & 652  & 3 \\
  & 03-05 21:02:01 & 177 & 170 & 1638 & 10 \\
  \hline
  \multirow{3}{*}{12465} & 04-12 12:03:52 & 99 & 99 & 2805 & 28 & \multirow{3}{*}{642-886}\\% new variable
  & 04-13 11:57:34 & 121 & 111 & 806  & 7 \\
  & 04-14 11:51:03 & 170 & 110 & 1639 & 15 \\
  \hline
  \multirow{3}{*}{17590} & 04-12 12:06:53 & 85 & 82 & 887 & 11 & \multirow{3}{*}{960-1372}\\% new variable
  & 04-14 11:33:24 & 49 & 49 & 392  & 8 \\
  & 04-15 11:18:03 & 112 & 106 & 2694 & 25 \\
  \hline
  \multirow{3}{*}{20788} & 04-26 11:34:25 & 63 & 63 & 687 & 11 & \multirow{3}{*}{905-1455}\\% new variable
  & 04-27 11:29:55 & 69 & 56 & 219  & 4 \\
  & 04-28 11:30:27 & 84 & 51 & 151 & 3 \\
  \noalign{\smallskip}\hline
\end{tabular}
\end{center}
\end{table}

Table 2 indicates that the measuring system of DRLR could track and measure space debris or defunct satellites with the size (RCS) of larger than 1m$^2$, the orbital altitude of less than 1000 km continuously at SHAO. The maximum distance is close to 1600 km. The sampling rate of laser ranging data could reflect the external characteristics of space targets.

\section{Orbit determination (OD) and residual analysis by DRLR}
\label{sect:Orbit}
Because the laser ranging data of DRLR is collected from the single stations, most of targets only have 1 arc per day except NORADID 20453. To meet the need of requirements of orbit determination, the 2 or 3 arcs are needed at least. In data processing, the non-cooperative targets are regarded as the particle and the size and centroid correction of targets are not considered (Huang 2003). The dynamic models used for the orbit determination are shown in Table 3.

\begin{table}
\begin{center}
\begin{minipage}[]{120mm}
\caption[]{Selection and description of dynamic model adopted in orbit determination}\label{Tab:publ-works}
\end{minipage}
\setlength{\tabcolsep}{1pt}
\small
 \begin{tabular}{cc}
  \hline\noalign{\smallskip}
  Perturbation Correction & Description \\
  \hline\noalign{\smallskip}
  Earth gravity & GGM02C(150*150) \\
  N-body perturbation & JPL DE 405 Ephemeris, including solar, lunar and planetary gravitations\\
  Solid earth tides / Pole tides & recommended model/ IERS2003 \\
  Ocean tides & TOPEX3.0 \\
  earth rotation parameter (ERP) & IERS EOP series \\
  Atmospheric drag & DTM94 \\
  Radiation pressure & Solar radiation pressure,Earth radiation pressure \\
  Relativistic perturbation & recommended model/ IERS2003 \\
  Tropospheric correction & MARINI/MURRAY model \\
  others & Weighted least squares method(Gauss-Jackson multistep method) \\
  \hline
\end{tabular}
\end{center}
\end{table}

All forces acting on the satellite are regarded as perturbations except the central gravitational force. Considering the un-modeled errors and observational errors, some used dynamic parameters are regarded as unknown parameters and estimated together with the orbit parameters (Tapley 1989). Based these consideration, the results of orbit residual of the mentioned six targets are shown in Table 4. The results include the estimated atmospheric drag parameter (Cd0) that is estimated in whole arc. Figures 1-6 indicate the orbit residuals of the 6 targets. The mean residual of the 6 targets are about 111cm. The maximum value is about 228cm and the minimum value is about 39cm.

\begin{table}
\begin{center}
\begin{minipage}[]{135mm}
\caption[]{OD of space debris with laser ranging data collected from single station (estimated Cd0)}\label{Tab:publ-works}
\end{minipage}
\setlength{\tabcolsep}{1pt}
\small
 \begin{tabular}{ccccc}
  \hline\noalign{\smallskip}
  NORADID & Start Time (UTC) & OD arc ($day$) & RMS ($m$) & Data Used (\%) \\
  \hline\noalign{\smallskip}
  \multirow{4}{*}{20453} & 03-04 11:49:08 & \multirow{4}{*}{2.1} & \multirow{4}{*}{1.10} & \multirow{4}{*}{97.9} \\% new variable
   & 03-05 10:38:06 & & & \\
   & 03-05 12:21:57 & & & \\
   & 03-06 11:12:27 & & & \\
  \hline
  \multirow{3}{*}{28222} & 03-02 21:06:44 & \multirow{3}{*}{2.1} & \multirow{3}{*}{0.39} & \multirow{3}{*}{100} \\% new variable
   & 03-03 21:06:39 & & & \\
   & 03-04 21:07:55 & & & \\
  \hline
  \multirow{3}{*}{11574} & 03-02 21:22:30 & \multirow{3}{*}{3.0} & \multirow{3}{*}{1.19} & \multirow{3}{*}{97.0} \\% new variable
   & 03-03 20:40:30 & & & \\
   & 03-05 21:02:01 & & & \\
  \hline
  \multirow{3}{*}{12465} & 04-12 12:03:52 & \multirow{3}{*}{2.1} & \multirow{3}{*}{0.67} & \multirow{3}{*}{99.8} \\% new variable
   & 04-13 11:57:34 & & & \\
   & 04-14 11:51:03 & & & \\
  \hline
  \multirow{3}{*}{17590} & 04-12 12:06:53 & \multirow{3}{*}{3.0} & \multirow{3}{*}{2.28} & \multirow{3}{*}{99.9} \\% new variable
   & 04-14 11:33:24 & & & \\
   & 04-15 11:18:03 & & & \\
  \hline
  \multirow{3}{*}{20788} & 04-26 11:34:25 & \multirow{3}{*}{2.1} & \multirow{3}{*}{1.01} & \multirow{3}{*}{100} \\% new variable
   & 04-27 11:29:55 & & & \\
   & 04-28 11:30:27 & & & \\
  \hline
\end{tabular}
\end{center}
\end{table}

\begin{figure}
  \centering
  \includegraphics[width=12.0cm, angle=0]{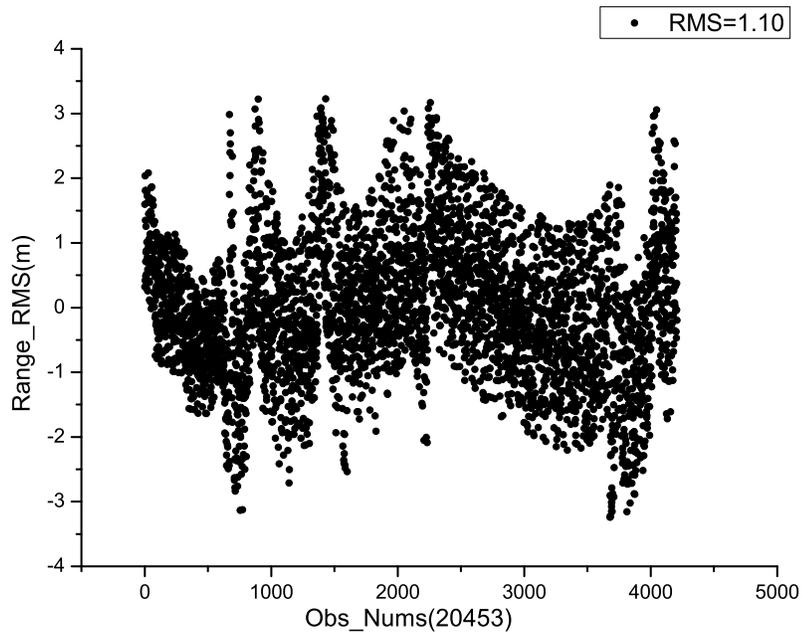}
  \begin{minipage}[]{40mm}
  \caption{Residuals of 20453}
  \end{minipage}
  \label{Fig1}
\end{figure}
\begin{figure}
  \centering
\includegraphics[width=12.0cm, angle=0]{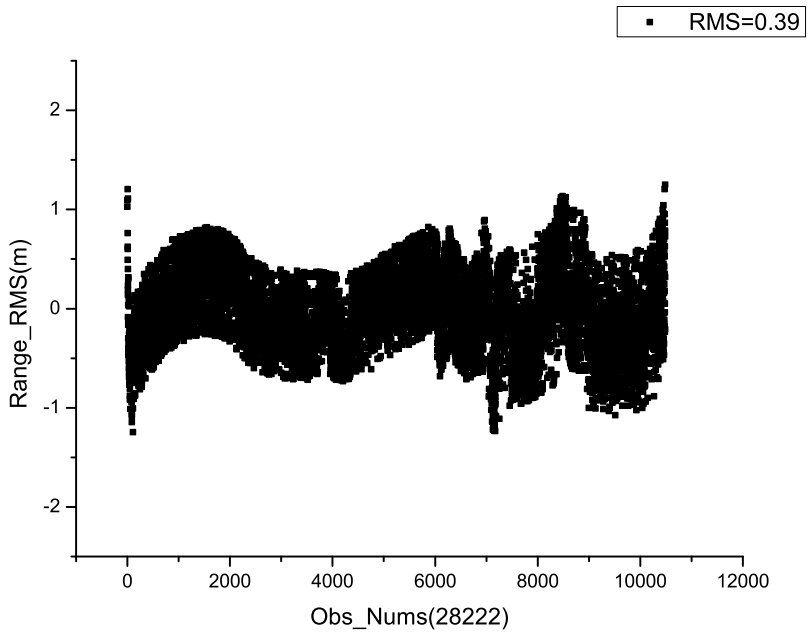}
  \begin{minipage}[]{40mm}
   \caption{Residuals of 28222}
  \end{minipage}
  \label{Fig2}
\end{figure}
\begin{figure}
  \centering
  \includegraphics[width=12.0cm, angle=0]{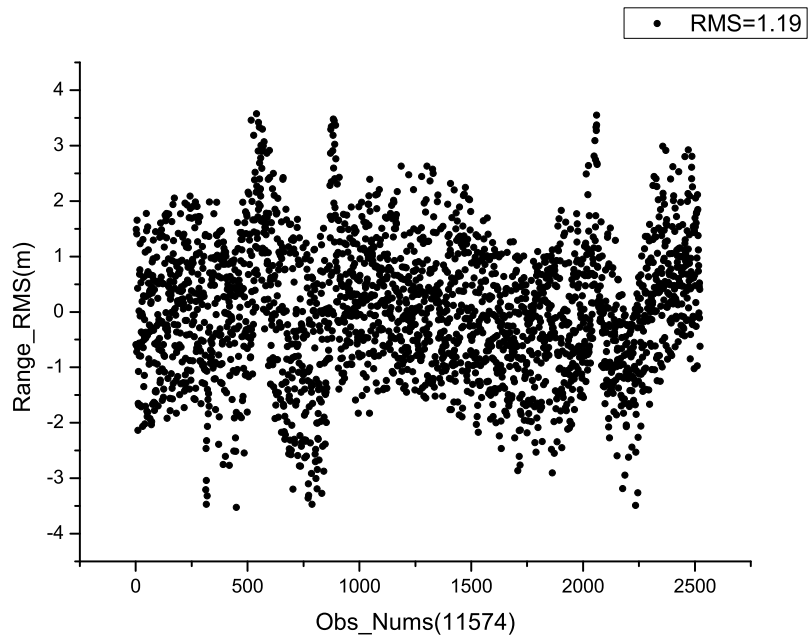}
  \begin{minipage}[]{40mm}
   \caption{Residuals of 11574}
  \end{minipage}
  \label{Fig3}
\end{figure}
\begin{figure}
  \centering
  \includegraphics[width=12.0cm, angle=0]{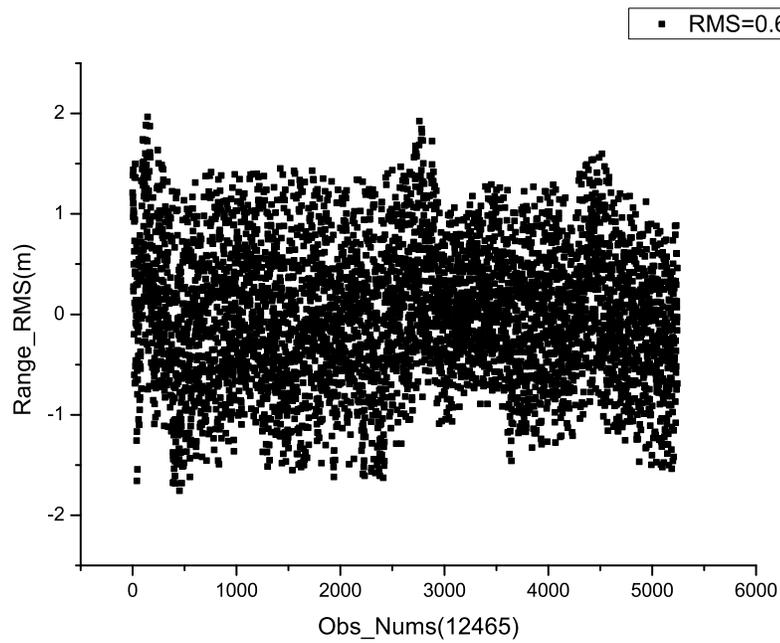}
  \begin{minipage}[]{40mm}
   \caption{Residuals of 12465}
  \end{minipage}
  \label{Fig4}
\end{figure}
\begin{figure}
  \centering
  \includegraphics[width=12.0cm, angle=0]{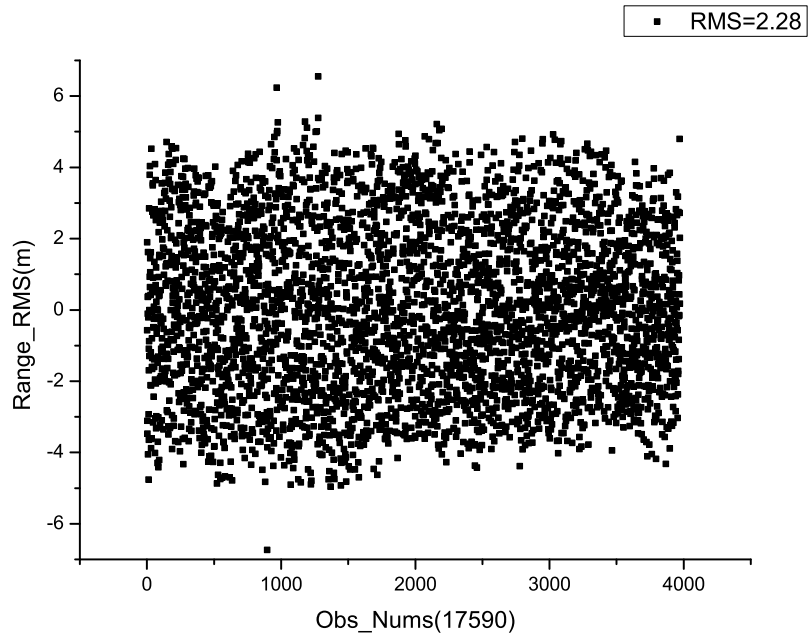}
  \begin{minipage}[]{40mm}
   \caption{Residuals of 17590}
  \end{minipage}
  \label{Fig5}
\end{figure}
\begin{figure}
  \centering
  \includegraphics[width=12.0cm, angle=0]{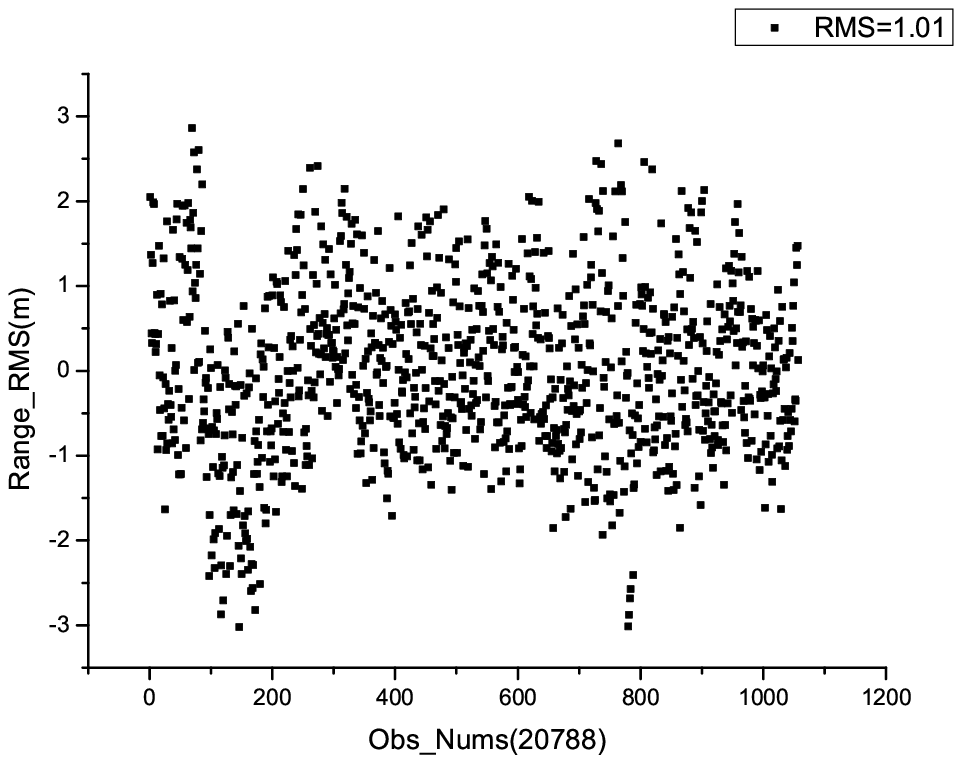}
  \begin{minipage}[]{40mm}
   \caption{Residuals of 20788}
  \end{minipage}
  \label{Fig6}
\end{figure}
Due to the size, attitude in orbit and other factors of non-cooperative space target, especially the space debris from the rocket wreckage, they generally fly with irregular flip. The OD residuals of these targets are from 39cm to 288cm which of accuracy is a little worse than that of the single laser ranging data (68-83cm)(SHAO 2009). The partial data of 20453(the starting time of arc is 2013-03-05 10:38:06, UTC) shown in Figure1 exhibits obviously periodic variation. When the high-frequency component (scattered points denote original results, the solid line denote filtered results) is filtered, the residuals computed and shown in Figure 7.

\begin{figure}
  \centering
  \includegraphics[width=12.0cm, angle=0]{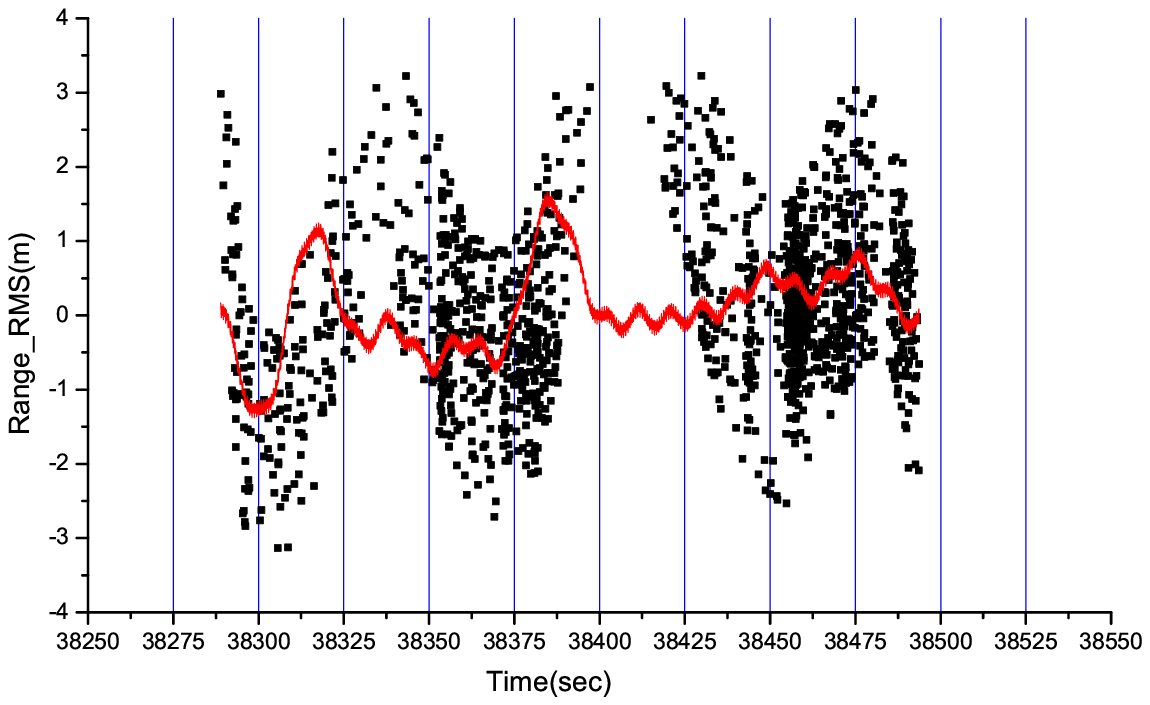}
  \begin{minipage}[]{100mm}
   \caption{Orbit residual of target 20453 after through low-pass filter}
  \end{minipage}
  \label{Fig7}
\end{figure}

\section{Evaluation of precision of OD from DRLR}
\label{sect:Eva}
For non-cooperative targets and space debris, under normal circumstances, it is difficult to obtain the accurate orbit product used to evaluate orbit result. Thus, the TLE (Two-Line Element set) provided on Space-Track website are used to test the reliability of the orbit results estimated with laser ranging data from a single station. After that, the method of orbit segmentation and orbital overlaps (Tapley 2004) is used to assess the accuracy of the orbit results of the targets which have adequate observations.

The target 20453 has 4 arcs in 3 days so that it has more adequate data and could meet the requirement of the assessing method of orbit segmentation and orbital overlaps (The same dynamic model and POD strategy is used). So the results from this target are used for assessing the accuracy of orbit determination. The results are shown in Table 5 and the plot of the deviation distribution of orbit is illustrated in figure 8.

\begin{table}
\begin{center}
\begin{minipage}[]{60mm}
\caption[]{The results of orbital overlaps}\label{Tab:publ-works}
\end{minipage}
\setlength{\tabcolsep}{1pt}
\small
 \begin{tabular}{ccccccccc}
  \hline\noalign{\smallskip}
  \multirow{2}{*}{NO.} & \multirow{2}{*}{Start Time(UTC)} & \multirow{2}{*}{Arcs}  & \multirow{2}{*}{OD arc(day)} & \multirow{2}{*}{Overlaps arc(day)} & \multicolumn{4}{c}{Deviation RMS:m} \\
   & & & & & R & T & N & 3D \\
  \hline\noalign{\smallskip}
  1 & 03-04 11:49:08 & 3 & 1.5 & \multirow{2}{*}{1.0} & \multirow{2}{*}{7.13} & \multirow{2}{*}{31.16} & \multirow{2}{*}{17.41} & \multirow{2}{*}{36.41}\\
  2 & 03-05 10:38:06 & 3 & 1.5 & & & & & \\			
   \hline\noalign{\smallskip}
\end{tabular}
\end{center}
\end{table}

\begin{figure}
  \centering
  \includegraphics[width=12.0cm, angle=0]{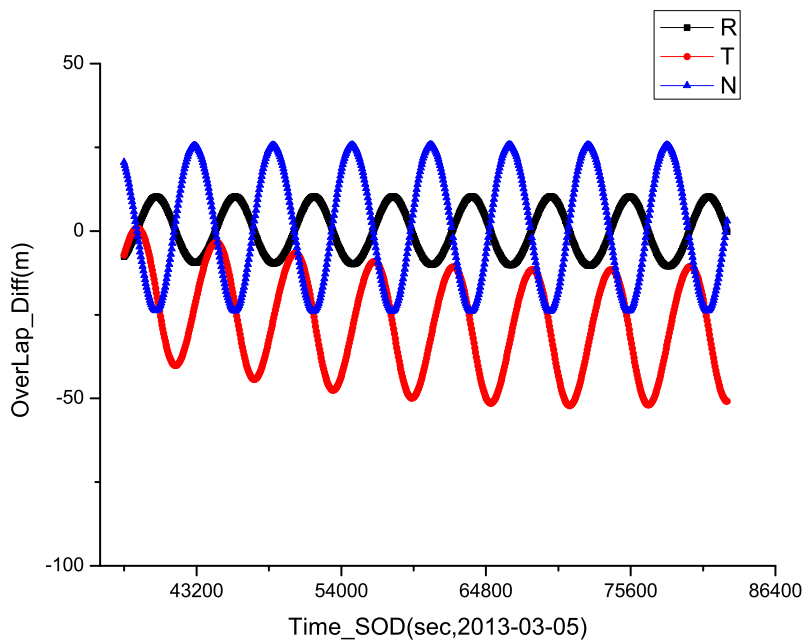}
  \begin{minipage}[]{125mm}
   \caption{The orbit deviation of target 20453 through orbit segmentation and orbital overlaps}
  \end{minipage}
  \label{Fig8}
\end{figure}

The results of orbit deviation between the whole arc and sub-arc are show in Table 6.
\begin{table}
\begin{center}
\begin{minipage}[]{90mm}
\caption[]{The deviation between the whole arc and sub-arc}\label{Tab:publ-works}
\end{minipage}
\setlength{\tabcolsep}{1pt}
\small
 \begin{tabular}{ccccccccc}
  \hline\noalign{\smallskip}
  \multirow{2}{*}{NO.} & \multirow{2}{*}{Start Time(UTC)} & \multirow{2}{*}{Arcs}  & \multirow{2}{*}{OD arc(sec)} & \multirow{2}{*}{Overlaps arc(day)} & \multicolumn{4}{c}{Deviation RMS:m} \\
   & & & & & R & T & N & 3D \\
  \hline\noalign{\smallskip}
  1 & 03-04 11:49:08 & 3 & 77+209+101 & 1.5 & 11.72 & 39.45 & 17.98 & 44.92\\
  2 & 03-05 10:38:06 & 3 & 209+101+131 & 1.5 & 4.64 & 12.39 & 0.96 & 13.27 \\			
   \hline\noalign{\smallskip}
\end{tabular}
\end{center}
\end{table}

From the results in Table 5 and Table 6, it can be seen that the accuracy of orbit determination of space debris is better than 50m, when the adequate laser ranging data from single station is processed. This result is better than that of microwave radar or optical-electric measurement system under the same conditions.

\section{Conclusions and discussions}
\label{sect:Conclusions}
Based on the technology of laser ranging already has, the laser measurement system was upgraded at SHAO. Some experiments on the space debris tracking are carried out and some good results are obtained. In these experiments, the collected data from some debris is processed and analyzed. The conclusions are as following:

1. The residual of OD shows that the mean precision of orbit is about 111cm. the maximum is about 228cm and the minimum is about 39cm;

2. The data from 18 targets and 43 effective arcs of laser ranging is processed and analyzed. The DRLR system could continuously track the large size (RCS$>$1m$^2$) of space debris at the orbit altitude of less than 1000km and with the maximum ranging close to 1600km.

3. By using the multi arcs of laser ranging data from single station, the orbit determination of space debris could be performed and the accuracy of orbit determination is better than 50m (orbital overlaps) with adequate data.

As new means of observing non-cooperative targets or space debris, the DRLR technology has significantly improved detection accuracy about 1m which is much higher than other space debris measurement equipment at the precision of tens of meters. Within ability of detection, DRLR system has gotten continuous targets measurements, and verified the feasibility of DRLR technology in the measurement to space debris with the high success rate. As the development of the high-precision measurement technology, DRLR system could provide high-precision data for orbit determination and collision warning in the future.

\begin{acknowledgements}
This work was funded by the National Natural Science Foundation of China (NSFC)under No.K0103110 and No.11173049.Thank Space-Track for providing debris' TLE of DRLR experiment.
\end{acknowledgements}

\label{lastpage}

\end{document}